%% file: main.tex
\documentclass[reprint,amsmath,amssymb,aps,prb,groupedaddress,nofootinbib,twocolumn,superscriptaddress]{revtex4-2}
\usepackage[english]{babel}
\usepackage[utf8]{inputenc}
\usepackage[colorinlistoftodos, color=green!40, prependcaption]{todonotes}

\input{preamble}

\begin{document}

\preprint{APS/123-QED}

\title{Altermagnetism from interaction-driven itinerant magnetism}

\author{Samuele Giuli}
    \email[Correspondence email address: ]{sgiuli@sissa.it}
    \affiliation{International School for Advanced Studies (SISSA), via Bonomea 265, 34136 Trieste, Italy}

\author{Carlos Mejuto-Zaera}
    \affiliation{International School for Advanced Studies (SISSA), via Bonomea 265, 34136 Trieste, Italy}

\author{Massimo Capone}
    \affiliation{International School for Advanced Studies (SISSA), via Bonomea 265, 34136 Trieste, Italy}
    \affiliation{CNR-IOM Democritos, Via Bonomea 265, 34136 Trieste, Italy}

\date{\today} 

\begin{abstract}
Altermagnetism, a new phase of collinear spin-order sharing similarities with antiferromagnets and ferromagnets, has introduced a new guiding principle for spintronic/thermoelectric applications due to its direction-dependent magnetic properties.
Fulfilling the promise to exploit altermagnetism for device design  depends on identifying materials with tuneable transport properties. The search for intrinsic altermagnets has so far focused on the role of anisotropy in the crystallographic symmetries and in the bandstructure. 
Here, we present a different mechanism  that approaches this goal by leveraging the interplay between a  Hubbard local repulsion and the itinerant magnetism given by the presence of van Hove singularities. We show that  altermagnetism is stable for a broad range of interactions and dopings and we focus on tunability of  the spin-charge conversion ratio.
\end{abstract}

\keywords{add keywords}

\maketitle

\input{sections/1_Introduction}  
\input{sections/2_Model}  
\input{sections/3_Method} 
\input{sections/4_Results}  
\input{sections/5_Conclusions}  

\paragraph*{Acknowledgements} \label{sec:acknowledgements}
We are grateful to C. Autieri, G. Bellomia, M. Fabrizio, R. Fernandes, M. Ferraretto, E. Linnér, J. Skolimowski, and A. Toschi
for helpful discussions. This work has been supported by 
National Recovery and Resilience Plan (NRRP) MUR Project  No. CN00000013-ICSC and PE0000023-NQSTI and by  MUR via PRIN 2020 (Prot. 2020JLZ52N 002) programs, PRIN 2022 (Prot. 20228YCYY7).

\bibliography{biblio, CMZ_refs}

\begin{widetext}

\appendix

\input{sections/S1.tex}
\input{sections/S2.tex}
\input{sections/S3.tex}
\input{sections/S4.tex}

\input{sections/S5.tex}

\end{widetext}

\end{document}

%% file: preamble.tex
\usepackage{amsthm}
\usepackage{mathtools}
\usepackage{physics}
\usepackage{xcolor}
\usepackage{graphicx}
\usepackage[left=23mm,right=13mm,top=35mm,columnsep=15pt]{geometry} 
\usepackage{adjustbox}
\usepackage{placeins}
\usepackage[T1]{fontenc}
\usepackage{lipsum}
\usepackage{csquotes}
\usepackage[pdftex, pdftitle={Article}, pdfauthor={Author}]{hyperref} 
\usepackage{caption}
\usepackage{bm}
\usepackage[caption=false]{subfig}

\usepackage[ruled]{algorithm2e}
\newcommand{\dagga}{{\phantom{\dagger}}}

\newenvironment{eqs}%
{\begin{equation} \begin{aligned}}%
{\end{aligned} \end{equation} }
\newcommand{\beal}{\begin{eqs}}
\newcommand{\eal}{\end{eqs}}

%% file: sections/1_Introduction.tex
\paragraph*{Introduction.} \label{sec:introduction}

The landscape of magnetic systems has been thrilled by the proposal and experimental realization of altermagnetism,  a new type of collinear spin-order which shares similarities with both antiferromagnets and ferromagnets~\cite{PhysRevX.12.031042,PhysRevX.12.040501}, while featuring new properties that hold a huge potential for fundamental science and applications.
Altermagnets are characterized by a zero net magnetization in the non-relativistic limit, similarly to antiferromagnets, but their electronic bands do not form a Kramer doublet, having instead a momentum-dependent spin-splitting. Importantly, the magnetic polarization of an altermagnet has a long-range order with d-wave (or higher, g- or i-wave) symmetry characterized by nodes. 
The applicative potential of altermagnets relies mostly on the  anisotropic magnetic properties  which have been proposed to 
give rise to giant magnetoresistance (GMR)
~\cite{PhysRevX.12.031042}, giant tunnel magnetoresistance
\cite{Shao2021SpinneutralCF}, non-vanishing spin splitting torque~ \cite{bai2022observation,gonzalez2021efficient} 
as well as anomalous Hall conductivity~\cite{Sattigeri2023,Fakhredine2023}. 

Despite the impressive and rapidly increasing body of work introducing candidate materials for altermagnetic ordering \cite{Cuono2023,Grzybowski2024,ferrari2024altermagnetismshastrysutherlandlattice,regmi2024altermagnetismlayeredintercalatedtransition,gonzález2024altermagnetismdimensionalcaruoperovskite,chang2024mysteriousmagneticgroundstate,parfenov2024pushingaltermagnetultimate2d,PhysRevB.110.L100402,PhysRevLett.132.176701,Milivojević_2024,mazin2023inducedmonolayeraltermagnetismmnpsse3,galindezruales2024altermagnetismhoppingregime,delre2024diracpointstopologicalphases,yu2024altermagnetismcoincidentvanhove,antonenko2024mirrorchernbandsweyl}, there are still few experimental evidences of this phase of matter. This is in part due to the lack of spin-resolved measurements and the small non-relativistic spin-splitting, which poses a challenge to its experimental resolution.
For $\text{RuO}_2$ films various altermagnetic phases have been realised~\cite{jeong2024} while the debate for bulk $\text{RuO}_2$~\cite{lin2024,kessler2024} is still open.
Compelling evidence for altermagnetic ordering has been presented instead for MnTe~\cite{PhysRevB.109.115102} and CrSb~\cite{reimers2024direct}.
Recently, it has been proposed that this direction-dependent magnetic phase could also arise from anisotropic local orbital states \cite{PhysRevLett.132.236701} where interactions can trigger the presence of an altermagnetic phase with the simultaneous realisation of orbital and magnetic ordering.

\begin{figure}[h!]
    \centering
    \includegraphics[width=\linewidth]{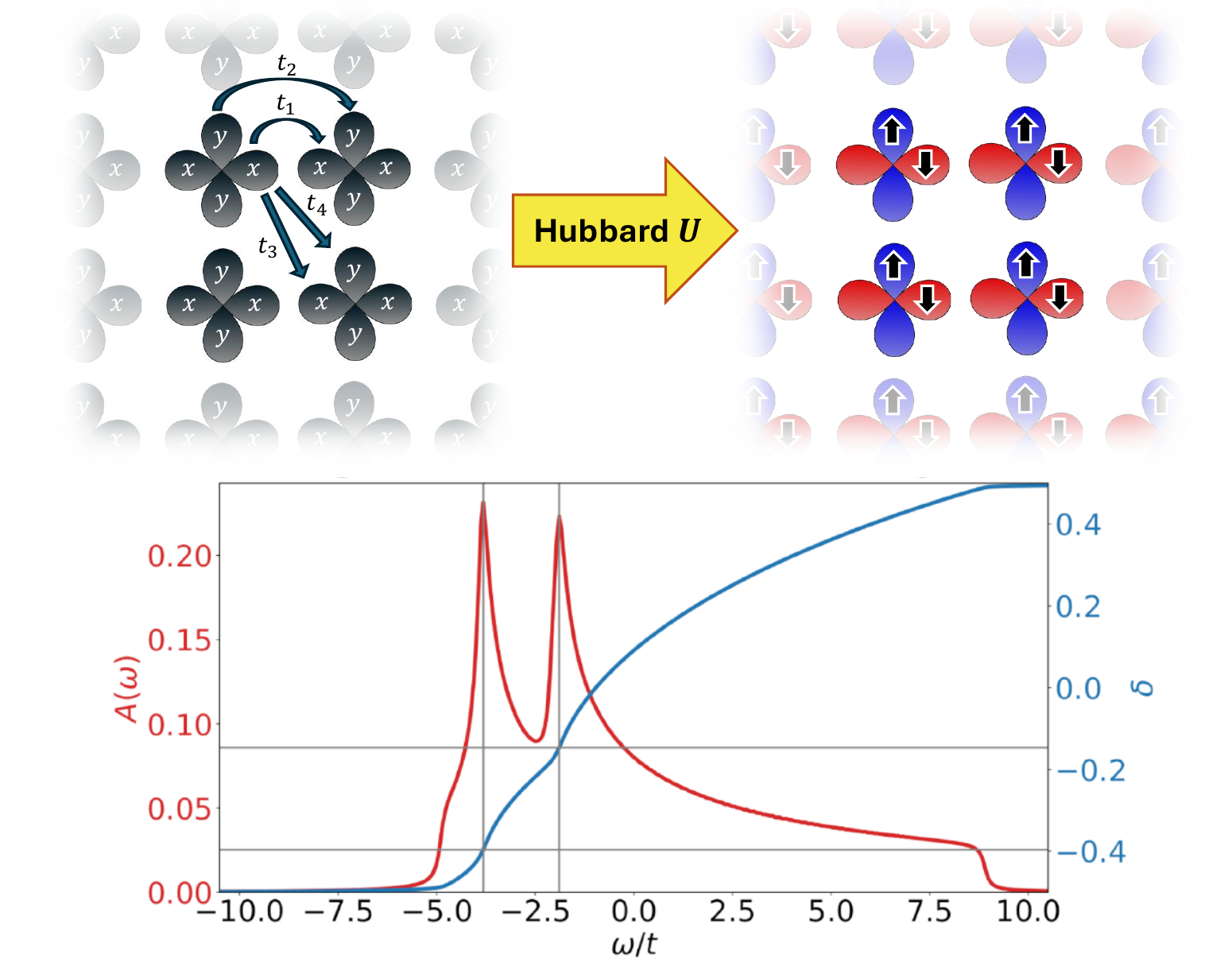}
    \caption{ Illustration of the interaction-driven phase transition from paramagnetic (top left) to altermagnetic (top right) and density of state of the non-interacting model (bottom-left axis) with doping (bottom-right axis) of the van Hove singularities in the non-interacting limit. }
    \captionsetup{justification=justified}  
    \label{fig:ALM_pic} 
\end{figure}

In the search for intrinsic altermagnetic materials, most proposals have leveraged chemical and structural manipulation to tailor the electronic structure. 
Yet, recently an alternative approach to let altermagnetism arise from interactions has gained traction~\cite{PhysRevLett.132.236701,ferrari2024altermagnetismshastrysutherlandlattice,delre2024diracpointstopologicalphases}.
This direction offers a particularly attractive perspective:  devising platforms with {\it tunable} altermagnetic properties exploiting strong electronic correlations, which characterizes materials with narrow bands and comparatively large values of the screened Coulomb repulsion.

Indeed, the two traditional collinear magnetic phases, ferromagnetism and antiferromagnetism, are linked to strong correlation physics.
Antiferromagnetism is an ubiquitous signature of strong correlations in undoped materials, while itinerant ferromagnetism is naturally connected with doped correlated materials.

The presence of ferromagnetism in the Hubbard model has been one of the driving forces behind the very introduction of the model. According to Nagaoka's theorem, it is the groundstate for a single hole and large interactions~\cite{PhysRev.147.392}, and it has been found in finite interactions~\cite{PhysRevB.31.4403,PhysRevB.35.3359,PhysRevB.77.220503,ulmke1998ferromagnetism} and for frustrated systems~\cite{PhysRevResearch.5.L022048,lebrat2024observation} as well.
In the triangular lattice Hubbard model the presence of a Van Hove singularity (vHs) at filling $n=3/4$ appears to be fundamental for the formation of itinerant ferromagnetism~\cite{xu2023frustration} and also in the $t-t'$ square lattice Hubbard model~\cite{PhysRevB.35.3359,PhysRevLett.78.1343}.
In twisted bilayer graphene, the presence of a vHs near half-filling has also been linked to $s$-wave and $s\pm$-wave Pomeranchuk instabilities~\cite{PhysRevB.102.125120,PhysRevB.102.125141}

In this work, building on the above arguments,  we propose a new paradigm where electronic correlations lead to the  onset of altermagnetic ordering.
Our proposal is based on a two-orbital model with  anisotropic orbitals, and a
van Hove singularity at moderate doping (See Figure \ref{fig:ALM_pic}).
We show that the altermagnetic ordering naturally emerges as a result of the cooperation between the correlation-driven itinerant magnetism and the $C_{4z}$ symmetry of inequivalent orbitals with a uniform, on-site, non-relativistic spin-splitting.
This mechanism relies on intrinsic correlation effects due to a local Hubbard repulsion that is not devised to explicitly break the spin and orbital symmetries. For this reason it can be realized in generic lattices even when spatial ordering is frustrated.
These features are indeed quite generic and could inspire the ab-initio and experimental search for altermagnetism candidates in unexplored classes of materials.

%% file: sections/2_Model.tex
\paragraph*{Model.}

We focus on a model introduced soon after the discovery of superconductivity in iron-based materials~\cite{PhysRevB.77.220503} as a minimal theoretical description. This model features two degenerate orbitals of $d_{xz}$ and $d_{yz}$ symmetry on each lattice site and local interactions parameterized by the intraorbital Hubbard $U$.
The one-body Hamiltonian on a square lattice reads:
\begin{equation}
    \hat{H}_0 = \sum_{\mathbf{k},s} 
    \begin{pmatrix} c^\dagger_{\mathbf{k} \textit{x} s} & c^\dagger_{\mathbf{k} \textit{y} s} \end{pmatrix}  
    \begin{pmatrix}  \ \epsilon_x(\mathbf{k}) & \epsilon_{xy}(\mathbf{k}) \\ \epsilon_{xy}(\mathbf{k})  & \ \epsilon_{y} (\mathbf{k}) \end{pmatrix}
    \begin{pmatrix} c_{\mathbf{k} \textit{x} s} \\ c_{\mathbf{k} \textit{y} s}     \end{pmatrix}    
    \label{eq:model}
\end{equation}
with
\begin{align}
    \epsilon_x(\mathbf{k}) =& -2t_1\cos k_x -2t_2 \cos k_y -4 t_3 \cos k_x \cos k_y \nonumber \\
    \epsilon_y(\mathbf{k}) =& -2t_2\cos k_x -2t_1 \cos k_y -4 t_3 \cos k_x \cos k_y \nonumber \\
    \epsilon_{xy}(\mathbf{k}) =& -4 t_4 \sin k_x \sin k_y 
    \label{eq:dispersion}
\end{align}
We set $t_1=-t$, $t_2=-1.75t$, $t_3=-0.85t$, and $t_4=-0.65t$, following  Ref \cite{PhysRevLett.132.236701}. 
This choice of parameters results in two van Hove singularities (vHs) in the bare density of states at doping $\delta_{vH} \approx -0.145$ and $\delta_{vH,2} \approx -0.396$. The existence of these vHs, together with the structure of the hoppings reflecting the $d_{xz}$ and $d_{yz}$ character of the orbitals, turn out to be the key ingredient to drive the altermagnetic state. Therefore, our results do not rely in the specific values of parameters as long as a vHs at reasonably small doping is present.

We will indeed focus on the doping region around the first vHs since the effect of the interaction around the second singularity is strongly reduced by the extreme doping. We introduce a purely intra-orbital Hubbard interaction, which, as discussed above, favours a ferromagnetic state close to a vHs on the single band Hubbard model~\cite{PhysRev.147.392,PhysRevB.31.4403,PhysRevB.35.3359}. The local interaction Hamiltonian therefore reads:
\begin{equation}
    \hat{H}_{int} = U \sum_{i, \alpha= \textit{x,y}} (n_{i \alpha \uparrow}-\frac{1}{2}) (n_{i \alpha \downarrow}-\frac{1}{2})
    \label{eq:interaction}
\end{equation}
where $n_{i \alpha s}$ is the density operator at site $i$, orbital $\alpha$ and spin $s$.

%% file: sections/3_Method.tex
\paragraph*{Method.}

We solve this model using the ghost rotationally invariant slave boson (gRISB) approach with mean-field boson decoupling~\cite{Lanata2022}, which is a generalization of RISB~\cite{Lechermann2007,Isidori2009} and 
equivalent to the ghost Gutzwiller method~\cite{Lanata2017}.
This is a non-perturbative, variational Ansatz to capture strong electronic correlation in model as well as \emph{ab initio} calculations of solids and molecules~\cite{Lanata2017,guerci2019,frank2021,Lanata2022,Mejuto2023a,Guerci2023,Lee2023a,Lee2023b,Mejuto2024,Lee2024}.
Within its embedding formulation, gRISB represents an interacting systems in terms of a pair of quasiparticle (qp) and impurity Hamiltonians.
The latter captures local correlations exactly, while the former features interaction-renormalized hoppings and one-body potentials.
The parameters of these Hamiltonians are determined self-consistently by matching their local one-body density matrices $\langle c^\dagger_{\alpha}c_{\beta}\rangle$, which makes the method computationally inexpensive~\cite{Fabrizio2007,lanata2015,Mejuto2024}.
Thus, we can use gRISB to rapidly and thoroughly explore the phase diagram of the multi-orbital model in Eq.~\eqref{eq:model}-\eqref{eq:interaction}, obtaining faithful local order parameters from the impurity Hamiltonians, as well as non-local correlation functions from the band structure of the qp Hamiltonian.
The description of correlated behavior in terms of a band structure theory greatly simplifies the study of the altermagnetic phase and its properties, and is enabled by the introduction of auxiliary states in the qp Hamiltonian, the eponymous ghosts. These ghosts enable capturing incoherent high-energy spectral features, for instance Hubbard bands, on top of the low-energy quasiparticle excitations~\cite{Lanata2017,Mejuto2023a,Mejuto2024}.
Throughout the paper we fix the number of ghost orbitals per physical band to be $N_{ghosts}=2$.
The main equations of the gRISB formalism are summarized in the Supplementary Material (SM)~\cite{supp}.

%% file: sections/4_Results.tex
\paragraph*{Results.}

\begin{figure} [h!]
    \centering
    \includegraphics[width=\linewidth]{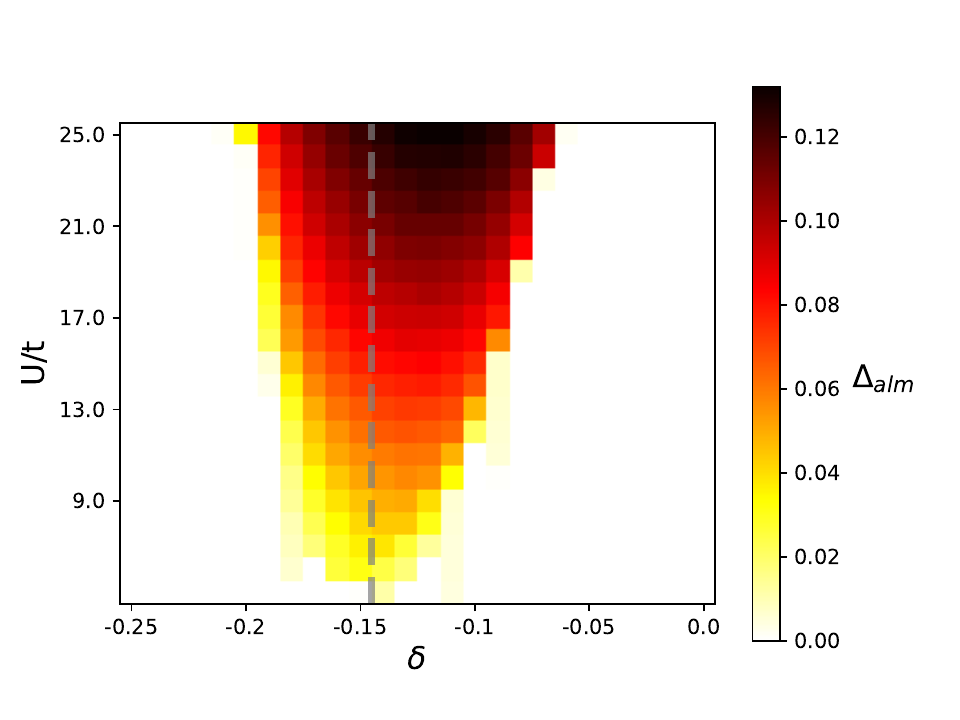}
    \caption{ Altermagnetic order parameter $\Delta_{alm}$ in Eq.~\eqref{eq:OP} as a function of the interaction $U$ and doping $\delta$. The gray dashed vertical line indicates the doping of the van Hove singularity in the non-interacting model.}
    \captionsetup{justification=justified}  
    \label{fig:phase-diag}
\end{figure}

\begin{figure*}
    \includegraphics[width=\textwidth]{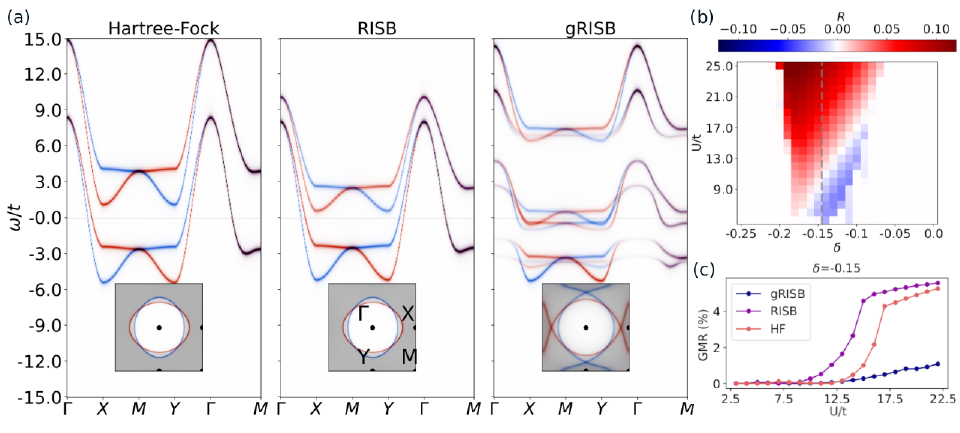}
    \caption{(a) Spin up (red) and spin down (blue) spectral functions along symmetry lines for $U/t=20$ and $\delta=-0.15$ using Hartree-Fock (left), RISB (center) and gRISB (right) with 2 ghosts. In the insets, the respective Fermi surfaces along the Brillouin zone, the shadowed area represent fermionic occupation: dark occupied, light empty. (b) Charge spin convertion ratio in Eq.~\eqref{eq:CSCratio} as a function of interaction ($U/t$) and doping ($\delta$). Positive values correspond to spin-current in the $[-110]$ direction while a negative ones are in the opposite direction. The gray dashed vertical line indicates the doping of the van Hove singularity in the non-interacting model. (c) GMR comparison between gRISB(dark), RISB(medium) and HF(light).}
    \label{fig:All_in_one}
\end{figure*}

We define the local altermagnetic order parameter as
\begin{equation}
    \Delta_{alm} = \frac{m_{x}-m_{y}}{2}
    \label{eq:OP}
\end{equation}
with $m_\alpha = n_{\alpha \uparrow}-n_{\alpha \downarrow}$ the spin magnetisation on orbital $\alpha = x,y$.
We present the ground state phase diagram in Figure \ref{fig:phase-diag}. The presence of the vHs at doping $ \delta_{vH} \approx -0.145$ (shown as a vertical, dashed line in Fig.~
\ref{fig:phase-diag}) is linked to the onset of spin ordering at large interaction $U$ with two orbitals having opposite magnetisation, realising the altermagnetic phase we propose in this paper.
In contrast to previously proposed interaction-driven altermagnetic phases~\cite{PhysRevLett.132.236701}, we note that the opposite magnetic moments related by $C_{4z}$ symmetry reside on \emph{the same atomic site}. 
The altermagnetic ordering is enhanced and it extends to a larger range of dopings with increasing $U/t$ while the maximum of the order parameter at fixed interaction is always found near $\delta = \delta_{vH}$.
Significant interactions result therefore in a large doping region where altermagnetism is stable and sizeable, avoiding the need for fine tuning the chemical potential, a particularly important point in view of applications to materials.
The altermagnetic phase is also characterized by the momentum-dependent Zeeman splitting shown in the gRISB bandstructure on the right side of Fig. \ref{fig:All_in_one}(a).

Our proposed altermagnetic state is energetically favorable with respect to a fully ferromagnetic one due to the presence of inter-orbital hoppings that, together with the local repulsion $U$, favor antialignment of the spins between the orbitals (see SM~\cite{supp}).
In addition, the presence of large next-nearest-neighbor hoppings ($t_3$ and $t_4$ in Eq.~\eqref{eq:dispersion}) frustrates bipartite ordering, therefore we expect the altermagnet to be stable with respect to other uniform or non-uniform orderings.
Further including inter-orbital interaction terms, that we neglected for the sake of simplicity, are not expected to destroy the altermagnetism as long the intra-orbital $U$ remains the largest energy scale, a condition which holds under very general circumstances.

As we mentioned above, transport properties are amongst the most appealing features of altermagnets. 
For instance, these phases allow for spin-splitter effect when applying an electric field in the $[110]$ direction.
Since $\hat{S}_z$ remains a good quantum number, we can compute the spin-independent conductivity for spin-up and spin-down:
\begin{equation}
    \sigma^{s}_{ab}(\omega) = -e^2 \frac{\langle \frac{1}{V_{BZ}}\sum_{\mathbf{k}} \partial_{k_a} \partial_{k_b} H^{s}_{\mathbf{k}} \rangle - \Lambda_{ab}^{s}(\omega) }{i( \omega + i 0^+)}
\end{equation}
with $s= \uparrow, \downarrow$, where $\Lambda_{ab}^{s}(\omega)$ is the $s$-spin current-current dynamical response that we compute as the elementary bubble without vertex corrections~\cite{PhysRevLett.65.243,PhysRevLett.68.2830}.

In the case of a static electric field in the $[110]$ direction the spin current $\mathbf{j}_s$ is orthogonal to the charge current $\mathbf{j}_c$~\cite{gonzalez2021efficient,PhysRevLett.132.236701}, hence we can use the Drude weight defined as 
\begin{equation}
    D^{s}_{ab} = \langle \frac{1}{V_{BZ}}\sum_{\mathbf{k}} \partial_{k_a} \partial_{k_b} H^s_{\mathbf{k}} \rangle - \Lambda_{ab}^{s}(\omega=0)
\end{equation}
to compute the charge-spin conversion ratio $R$ as
 \begin{equation}
     R = \frac{|\mathbf{j}_s|}{|\mathbf{j}_c|} = \frac{\sqrt{ \sum_{a} (j^\uparrow_a-j^\downarrow_a)^2} }{\sqrt{ \sum_{a} (j^\uparrow_a+j^\downarrow_a)^2} }
     \label{eq:CSCratio}
 \end{equation}
 where $j^s_{a} = \sum_\beta D^s_{a b} A_b$ and $A_b = A$ is the vector potential for $a,b= x,y$ directions.

We plot the ratio $R$ in Figure~\ref{fig:All_in_one}~(b) with a positive sign for spin current in the $[-110]$ direction and negative for the opposite one. We notice an enhancement of $R$ when increasing both doping and/or the interaction $U/t$. The ratio rises up to ${R\approx9\%}$ for a typical value of $U/t=20$, corresponding to an angle between the spin-up and spin-down currents of ${\theta\approx10^{\circ}}$. 
We also notice that the spin current changes sign along a straight line in our $U$-$\delta$ diagram.

The inversion of the spin current is related to the different  spin polarization of the carriers in $x$ and $y$ orbitals. The Drude weight is dominated by the contribution of the electrons at X in the Brillouin zone (See SM~\cite{supp}). The $X$ orbital hosts a majority of spin-up carriers while the $Y$ orbital a majority of spin-down carriers. 
This phenomenon opens the possibility of changing the sign of the spin current tuning the interaction via mechanical or chemical pressure.

In order to assess how relevant strong correlations are to this altermagnetic ordering, we plot the band structure for an interaction $U/t=20$ and doping $\delta=-0.15$ for Hartree-Fock (HF, left), RISB (center) and gRISB (right) in Figure~\ref{fig:All_in_one}~(a).
The HF and RISB result are qualitatively similar, with the latter showing a moderate renormalisation due to the local Hubbard interaction.
The gRISB band structure instead shows a stronger renormalisation and a large transfer of specral weight toward the lower and upper Hubbard bands.
Moreover, the effect of strong correlation clearly reduces the non-relativistic spin-splitting and, more importantly, also changes the topology of the Fermi surface as shown in the insets of Figure~\ref{fig:All_in_one}~(a) with only gRISB predicting the presence of pockets around $X$ and $Y$ points in the Brillouin zone at large $U/t$.
The presence of Dirac band crossing along the $\Gamma X$, $\Gamma Y$, $MX$ and $MY$ lines is present only in the gRISB calculations for strong interactions. The reason is the existence of the following critical condition on the intra-orbital magnetic gap ($\Delta_{mag}$) above which the Dirac points disappear (See SM and \cite{delre2024diracpointstopologicalphases}).
\begin{equation}
    \frac{\Delta_{mag}}{|t_1 - t_2|} \leq 1
\end{equation}
The reduction of the magnetic gap in gRISB is a key ingredient for the persistence of Dirac points.
As shown in pannel (c) of Figure~\ref{fig:All_in_one}, the giant magnetoresistance, defined as:
\begin{equation}
    GMR = \frac{1}{2}(\frac{\sigma_{xx}^{\uparrow}}{\sigma_{xx}^\downarrow}+\frac{\sigma_{xx}^{\downarrow}}{\sigma_{xx}^\uparrow} -2)
    \label{eq:GMR}
\end{equation}
is also dramatically reduced by the inclusion of the high-energy ghosts improving the description of dynamical correlation effects. 
We observe indeed a reduction of a factor of almost 5 when comparing gRISB to RISB in the large interaction limit.
This effect could partially resolve the mismatch between density functional theory predictions of ALM and the lack of spin-gap resolution in ARPES experiments.

%% file: sections/5_Conclusions.tex
\paragraph*{Conclusions and Outlook.}
\label{sec:conclusions}
We proposed a new microscopic mechanism for interaction-generated altermagnetism linked to itinerant magnetism in the vicinity of a simple van Hove singularity, and we reported numerical evidence for its realization in a  
two-orbital model whose non-interacting bandstructure was introduced to describe iron-based superconductors.
The altermagnetic state appears in a wide range of dopings and values of the interaction strength, presenting a maximum of the order parameter centered near the van Hove singularity $\delta_{vH}$.

We analyzed the charge-spin conversion ratio responsible for the spin-splitter effect of the altermagnetic state and we observed a change of the spin current direction as a function of doping and interaction. This phenomenon opens the intriguing possibility to control the direction of the spin-current tuning the interaction exerting mechanical or chemical pressure on candidate materials.

Finally we discussed the effect of strong correlations beyond mean-field in the theoretical description. These  clearly affect the nature of the altermagnetic state changing the shape of the band structure and the topology of the Fermi surface. As a result, they reduce the altermagnetic spin gap and giant magnetoresistance with respect to simple mean-field calculations (Hartree-Fock). This effect could justify the difficulties to experimentally find photo-emission evidence for altermagnetism in strongly correlated compounds.
We note that a related mechanism has been proposed for $\kappa$-Cl in Ref. \cite{yu2024altermagnetismcoincidentvanhove} which, unlike here, requires two coincident van Hove singularities at the M point to stabilize altermagnetic ordering.

Possible realisations of this mechanism could be found in Fe-based materials where evidence for altermagnetism  has been reported for Hematite ($\alpha-\text{Fe}_2\text{O}_3$)~\cite{galindezruales2024altermagnetismhoppingregime} and $\text{FeSe}$~\cite{mazin2023inducedmonolayeraltermagnetismmnpsse3}.
Further, systems where the t2g orbital physics is dominant, like $\text{Ca}_2\text{RuO}_4$~\cite{Cuono2023}, could be suitable platform when the t2g bands are partially filled.
Another promising platform are Moiré systems, where electronic properties can be tailored precisely and a wide palette of systems can be designed. In this context, evidence of itinerant ferromagnetism linked to the presence of a van Hove singularity~\cite{Tang2020SimulationOH,PhysRevB.107.235105,PhysRevB.109.045144} and numerical evidence of altermagnetic ordering~\cite{PhysRevMaterials.8.L051401,guo2024valleypolarizationtwistedaltermagnetism} have been reported. 

In general, we emphasize the three key ingredients of our mechanism:
(i) Two anisotropic orbitals related by a proper or improper rotation (e.g. in the model we present the $x$ and $y$ orbitals are related by $C_{4z}$ symmetry) (ii) a strong local Hubbard interaction $U$ and hoppings between the two orbitals to generate direct inter-orbital spin exchange and (iii) the presence of a vHs at an incommensurate filling promoting itinerant magnetic properties.

Moreover, frustration due to the lattice (e.g. triangular lattice) or due to hopping terms beyond nearest-neighbours  (e.g. $t_3$ in this paper) opposes to bipartite ordering, hence favoring our altermagnetic state. In addition, materials with the previous properties and an inverted Hund's coupling $J$ arising from electron-lattice coupling~\cite{Capone2002,Capone2009} may have an increased tendency to anti-align the spins on different orbitals, assisting the altermagnetic ordering we presented.
Conversely, a positive sign $J$ would only favor inter-orbital spin-alignment.
While for some materials, such as $\text{Ca}_2\text{RuO}_4$~\cite{Cuono2023}, it is reasonable to expect that our altermagnetic order will survive, beyond a certain value of $J$ the energetic gap between the spin-aligned and anti-aligned phases (See SM~\cite{supp}) should lead to its disappearance.

In this work, we have not considered the effect of spin-orbit coupling. We do not expect this to spoil our mechanism~\cite{gonzalez2021efficient}, but rather to enrich the phase diagram by breaking spin collinearity, thereby giving rise to non-trivial band topology~\cite{antonenko2024mirrorchernbandsweyl,sciadv.aaz8809}.

%% file: sections/S1.tex
\section{Conductivity and Drude weight}
In presence of a vector potential $A$ that is approximately constant at the lattice scale, the Peierls substitution consists in replacing the hopping integrals as follows:
\begin{equation}
    t_{ij} \rightarrow t_{ij} \exp{ -i \frac{e}{\hbar} \int_{\mathbf{R}_i}^{\mathbf{R}_j} \mathbf{A}( \mathbf{r},t ) \cdot d \mathbf{r} }
\end{equation}

We consider a vector potential in the $[110]$ direction, i.e., $(\mathbf{A})_x=(\mathbf{A})_y$ and $(\mathbf{A})_z =0$.
Expanding the kinetic terms of the full energy for small uniform vector potentials we find:
\begin{align}
    \hat{K} =& \sum_{ij,\alpha \beta} t^{\alpha \beta}_{ij} c^\dagger_{\mathbf{R}_i \alpha} c_{\mathbf{R}_j \beta} \{ 1 + i\frac{e}{\hbar} \mathbf{A}(t)\cdot (\mathbf{R}_i - \mathbf{R}_j )\nonumber \\
    -& \frac{1}{2}(\frac{e}{\hbar})^2 [\mathbf{A}(t)\cdot (\mathbf{R}_i - \mathbf{R}_j )]^2 + \mathcal{O}(\mathbf{A}^3) \}
\end{align}
with $\alpha,\beta=(\sigma,m)$ spin-orbital indexes. Replacing $\mathbf{R}_i = \mathbf{R}$ and $\mathbf{R}_i - \mathbf{R}_j = \mathbf{d}$
The total current in the $a=x,y$ direction is given by:
\begin{align}
    \hat{J}_a = \frac{\delta \hat{K}}{\delta A_a}  =&\sum_{\mathbf{R} \mathbf{d}} \sum_{\alpha \beta} t^{\alpha \beta}_{\mathbf{R},\mathbf{R}+\mathbf{d}} c^\dagger_{\mathbf{R}\alpha} c_{\mathbf{R}+\mathbf{d} \beta} \big[ -i\frac{e}{\hbar} d_a +\nonumber \\
    & (\frac{e}{\hbar})^2 d_a \mathbf{d}\cdot \mathbf{A}(t) \big]\\
    =& e[  \hat{J}^P_a + \hat{J}^D_a ]
\end{align}
with a paramagnetic contribution ($ \hat{J}^P_a $) independent from $\mathbf{A}$ and a diamagnetic contribution ($ \hat{J}^D_a $ ) that is $\mathbf{A}$ dependent. In units for which $c=\hbar=1$, one defines the conductivity $\sigma$ in linear response as:
\begin{equation}
    j_a (\omega) = \sigma_{ab}(\omega) \epsilon_b ( \omega )
\end{equation}
where $j_a = \langle \hat{J}_a \rangle$, and
\begin{equation}
    \mathbf{A}(\omega) = \frac{ \epsilon (\omega)}{i(\omega + i0^+)}.
\end{equation}
 From this, it follows that one can rewrite $j_a$ as
\begin{equation}
    j_a(\omega) \frac{1}{i(\omega + i0^+)} = \sigma_{ab} (\omega) A_b (\omega).
\end{equation}

Therefore the conductivity for spin $s$ can be computed from the contribution to first order in $\mathbf{A}$ to the total current as:
\begin{equation}
    \sigma^s_{ab} (\omega) = -e^2 \frac{\langle  \frac{1}{V_{BZ}} \sum_{\mathbf{k}}\partial_{k_a} \partial_{k_b} H^s (\mathbf{k}) \rangle - \Lambda^s_{ab}(\omega) }{i(\omega+i0^+)}
\end{equation}
with $\langle  \frac{1}{N_k} \sum_{\mathbf{k}}\partial_{k_a} \partial_{k_b} H^s (\mathbf{k}) \rangle$ the expectation value of the second derivative of the kinetic Hamiltonian for spin $s$ in the $a$ and $b$ directions in reciprocal space and $\Lambda^s_{ab}$ the uniform paramagnetic current ($\hat{J}^P$) susceptibility for spin $s$ that we compute as the solely elementary bubble, that is:

\begin{align}
    \Lambda_{ab}^s(\omega)=&\frac{1}{V_{BZ}} \sum_{\mathbf{k}} \int_{\omega_1} \int_{\omega_2} \frac{1}{\pi^2} \text{Tr}\Big[\Im{\hat{G}^s(\mathbf{k},\omega_1)}\hat{V}_a \nonumber \\
    &\Im{(\hat{G}^s)^T (\mathbf{k},\omega_2)} \hat{V}_b \Big] \frac{f_{FD}(\omega_1) - f_{FD}(\omega_2) }{\omega_1 - \omega_2 + \omega +i0^+}
\end{align}
with $\hat{G}$ the Green's function in the spinorial representation for the orbitals and $\hat{V}_a$ is the current vertex along direction $a$.
The Drude weight is the residue of the pole at $\omega=0$ of the conductivity:
\begin{equation}
    D^s_{ab} = -e^2 \big[ \langle  \frac{1}{V_{BZ}} \sum_{\mathbf{k}}\partial_{k_a} \partial_{k_b} H^s (\mathbf{k}) \rangle - \Re{ \Lambda^s_{ab}(\omega = 0)} \big]
\end{equation}

%% file: sections/S2.tex
\section{Current inversion}
In Figure \ref{fig:All_in_one} panel (b) we show that the spin current present an inversion along a line in the $U-\delta$ space. This inversion can be understood by looking at the contributions coming purely from orbital $x$ and $y$ to the diamagnetic component of the Drude weight, i.e., the Hessian of the non-interacting Hamiltonian in $k$-space. Namely we can compare the $\mathbf{k}$-space resolved contributions to the Drude weight along $x$ direction of the two orbitals for each spin $s$ and orbital $\alpha$, that are:
\begin{align}
    [\tilde{j}_{\alpha}^s (\mathbf{k}) ]_x   = {(-1)^s} \langle \partial_{k_x} \partial_{k_x} (H^s_{\mathbf{k}})_{\alpha \alpha} \rangle  
\end{align}

\begin{figure}[h!]
    \centering
    \includegraphics[width=\linewidth]{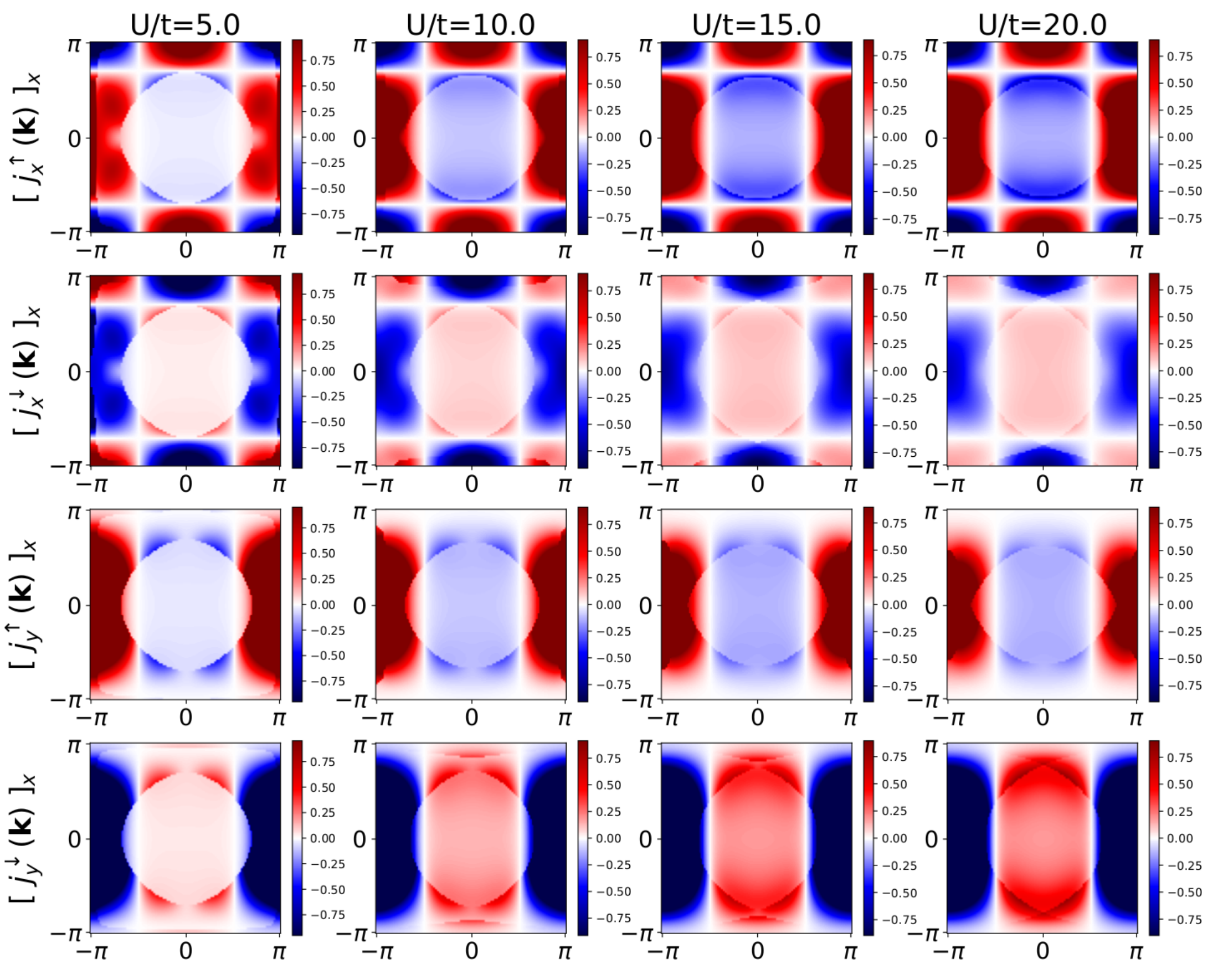}
    \caption{Momentum resolved contributions to the diamagnetic Drude weight, orbital and spin diagonal for doping $\delta=-0.13$. From left to right the interactions $U/t=5,10,15,20$. From top to bottom the spin-$\uparrow$ orbital X, spin-$\downarrow$ orbital X, spin-$\uparrow$ orbital Y and spin-$\downarrow$ orbital Y.}
    \label{fig:mom_orb_spin_resolved}
\end{figure}

\begin{figure}[h!]
    \centering
    \includegraphics[width=\linewidth]{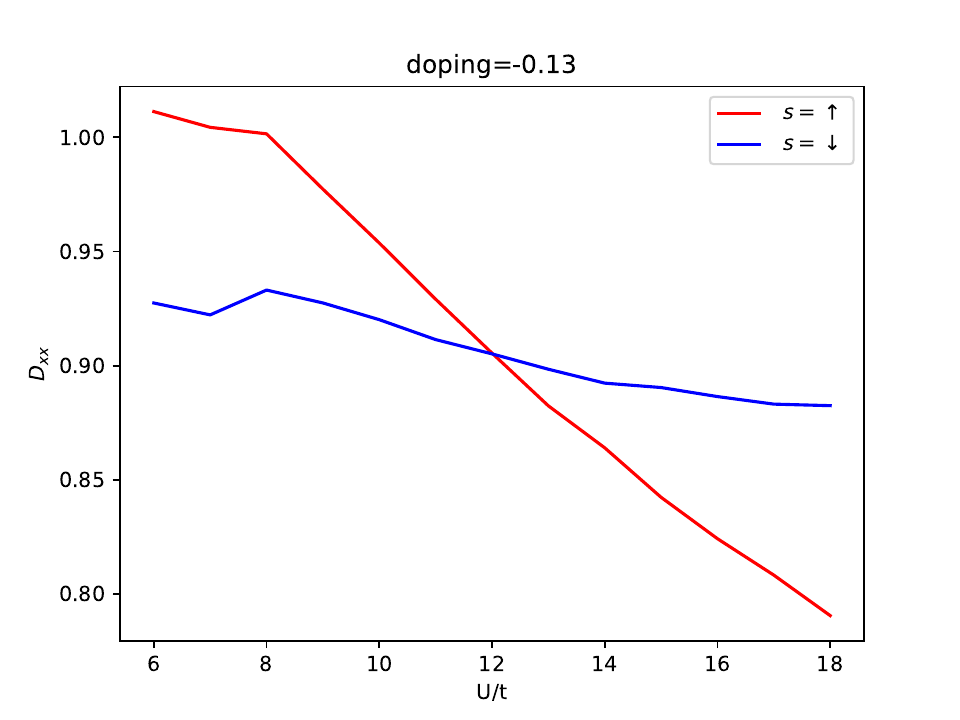}
    \caption{Orbital-diagonal contributions to the diamagnetic Drude weight for doping $\delta=-0.13$}
    \label{fig:Drude_spin_resolved}
\end{figure}

We show such contributions in Fig. \ref{fig:mom_orb_spin_resolved}. From left to right, increasing the interaction $U$ we see that for orbital X it is the up current that is strenghtened ($ [\tilde{j}_{x}^\uparrow (\mathbf{k}) ]_x > [\tilde{j}_{x}^\downarrow (\mathbf{k}) ]_x  $) while for orbital Y it is the down current ($ [\tilde{j}_{y}^\downarrow (\mathbf{k}) ]_x >  [\tilde{j}_{y}^\uparrow (\mathbf{k}) ]_x  $).
This means the two orbitals have spin-polarized currents, as to be expected since they are ferromagnetically ordered. This ordering is enhanced by increasing $U$. Moreover, we see that the biggest contribution to the spin current comes from the pockets found around the X point in the Brillouin zone.
If we look at the spin resolved, orbital-diagonal contributions to the diamagnetic Drude weight we can see that a crossing happens at a finite value of the interaction (See Fig. \ref{fig:Drude_spin_resolved}), or similarly a finite value of the doping due to a different slope of as a function of U for the currents associated to the two spin components.

The Drude weight for small $U/t$ is dominated by spin-$\uparrow$ electrons, belonging mostly to orbital X (See Figure \ref{fig:mom_orb_spin_resolved}) while increasing $U$ and crossing the spin-current inversion line the majority carriers become spin-$\downarrow$ electrons belonging to orbital Y. 
This phenomenon can be understood in terms of an effective Drude model for the pockets at the point X in the Brillouin zone.
For a free electron gas the Drude weight is $D = \frac{e^2 \pi n}{m}$ with $n$ the number and $m$ the mass of the electrons.
In our case the spin-$\uparrow$ and spin-$\downarrow$ electrons have different masses due to the different orbital nature. Moreover their chemical potentials are also different due to the magnetic ordering.
The number of spin-$s$ electrons for the pocket around the X point is proportional to the integral of the paraboloid centered in X. For a paraboloid of height $h$ and radius $r$ the volume is given by $(\pi/2)r^2 h$. In our case the height is given by the electrochemical potential $\mu_s$. The radius instead can be obtained from $(\mathbf{k}-\mathbf{k}_X)^2/(2 m_s) = -\mu_s  $ where $r^2=(\mathbf{k}-\mathbf{k}_X)^2=-2 m_s\mu_s$. Note that in the parabolic approximation $(\mathbf{k}-\mathbf{k}_X)^2$ is constant therefore $\mu_s \propto -1/m_s$. From this approximation we can say that the spin resolved Drude weight must be $D_s \propto -\mu_s / m_s$.
Doping the altermagnet in a rigid band approximation can be seen has a rigid shift of $\mu_s$ for both spin types. Doping with holes is a positive shift in $\mu_s$ therefore for a fixed change in doping the Drude weight for the two spins is increased proportionally to their inverse mass. The effective mass of the spin-$\downarrow$ electrons at the Fermi level is smaller and their contribution to the Drude weight decreases less than the spin-$\uparrow$ with a larger mass at the Fermi level. The two Drude weights as a function of doping therefore have different slopes and meet at a finite value of the doping.
The effect of the interaction $U$ instead is to increase the magnetic gap therefore increasing the splitting between $\mu_\uparrow$ and $\mu_\downarrow$. This is easily obtained at mean-field level, but it is expected to be a general effect.
Similarly, in a theory with fixed masses $m_s$ this will lead to an increase of $\mu_\downarrow$ with respect to $\mu_\uparrow$.

%% file: sections/S3.tex
\section{gRISB equations} 

In this appendix we succinctly summarize the main ingredients and the algorithmic structure of the gRISB method with mean-field treatment of the bosons.
As noted in the main body of the paper, this is equivalent to the ghost Gutzwiller approach.
For a more detailed discussion of the origin and heuristic meaning of the equations, as well as implementational details, we refer to previous work in the field~\cite{Lanata2017,Mejuto2023a, Mejuto2024} and references therein.

For simplicity, we shall assume that we are interested in a lattice Hamiltonian with space translation invariance of the form

\begin{equation}
    H_{phys} = \sum_{I} H^{loc}_I + H_{latt},
    \label{eq:Hphys}
\end{equation}

where capital indices $I$ run over lattice sites, $H^{loc}_I$ represents the local, possibly interacting, site Hamiltonian

\begin{equation}
    H^{loc}_I = \sum_{\alpha\beta}\ t^{II}_{\alpha\beta}\ c^\dagger_{I\alpha}c^\dagga_{I\beta} + \sum_{\alpha\beta\gamma\delta}\ U_{\alpha\beta;\gamma\delta}\ c^\dagger_{I\alpha}c^\dagger_{I\gamma}c^\dagga_{I\delta}c^\dagga_{I\beta}, \label{eq:Hloc}
\end{equation}

and $H_{latt}$ is the non-interacting lattice Hamiltonian, including the inter-site hopping terms

\begin{equation}
    H_{latt} = \sum_{I\neq J}\sum_{\alpha\beta}\ t^{IJ}_{\alpha\beta}\ c^\dagger_{I\alpha}c^\dagga_{J\beta} = \sum_{\boldsymbol{k}}\sum_{\alpha\beta}\epsilon_{\alpha\beta}(\boldsymbol{k})\ c^\dagger_{\boldsymbol{k}\alpha}c^\dagga_{\boldsymbol{k}\beta}.
    \label{eq:Hlatt}
\end{equation}

The orbitals of the physical Hamiltonian are indicated with Greek indices $\alpha,\beta,\gamma,\delta$, and may include the spin index.
In the last equality of Eq.~\eqref{eq:Hlatt} we have used the space translational invariance to write the lattice Hamiltonian in momentum space, with the band structure dispersion $\epsilon_{\alpha\beta}(\boldsymbol{k})$.
Within its embedding formulation, gRISB maps $H_{phys}$ into a quasiparticle Hamiltonian $H_{qp}$.
This is an effective, non-interacting Hamiltonian, in which correlation effects are represented in terms of band structure renormalizations $R$ and potentially local one-body potentials $\lambda$.
This leads to

\begin{equation}
    H_{qp} = \sum_{\boldsymbol{k}}\sum_{ab}\left[\sum_{\alpha\beta}R^\dagger_{a\alpha}\epsilon_{\alpha\beta}(\boldsymbol{k})R_{\beta b}-\lambda_{ab}\right]d^\dagger_{\boldsymbol{k}a}d^\dagga_{\boldsymbol{k}b},
    \label{eq:Hqp}
\end{equation}

where we use Latin indices to distinguish quasiparticle orbitals from physical ones, besides denoting the creation operators with the letter $d$ instead of $c$.
Additionally, we may choose to define $H_{qp}$ over an enlarged Hilbert space by including the so-called ghost orbitals to improve the description of strong correlation~\cite{Lanata2017,Mejuto2023a}.

The parameters $R$ and $\lambda$ in $H_{qp}$ are self-consistently determined with the solution of an impurity model $H_{imp}$, which brings in information about the local interactions in the system.
This impurity model contains the physical orbitals of a single site in $H_{phys}$, dressed by a non-interacting quasiparticle bath, such that

\begin{equation}
    H_{imp} = H^{loc}_0 + \sum_{\alpha a}\left(V_{a\alpha}\ d^\dagger_{a}c^\dagga_\alpha  + \mathrm{h.c.}\right) - \sum_{ab}\lambda^c_{ab}\ d^\dagga_b d^\dagger_a .
    \label{eq:Himp}
\end{equation}

Here, $H_{0}^{loc}$ denotes the local Hamiltonian contribution for some arbitrary site in the lattice (as all are equal by translational invariance).
The bath parameters $V$ (hybridizations) and $\lambda^c$ (local potential) are also determined self-consistently, by imposing that the ground state local one body reduced density matrices of $H_{qp}$ and $H_{imp}$ match

\begin{equation}
    \langle d^\dagga_b d^\dagger_a \rangle_{imp} \overset{!}{=} \langle d^\dagger_{0a} d^\dagga_{0b} \rangle_{qp}\equiv \Delta^{qp}_{ab}.
    \label{eq:gGutscf}
\end{equation}

Again, $d^\dagger_{0a}$ denotes orbital $a$ at an arbitrary lattice site.
This self-consistency can be achieved iteratively, starting form some guess $R$ and $\lambda$ parameters, as follows:

\begin{enumerate}
    \item The current $R$ and $\lambda$ define a $H_{qp}$ according to Eq.~\eqref{eq:Hqp}.
    From this, compute the ground state energy and local 1RDM $\Delta^{qp}_{ab} = \langle d^\dagger_{0a}d^\dagga_{0b}\rangle_{qp}$.
    \item Evaluate the impurity model parameters $V$ and $\lambda^c$.
    These can be computed from the $H_{qp}$ ground state.
    The hybridization $V$ follows
    \begin{equation}
        V = \frac{1}{V_{BZ}}\sqrt{\Delta^{qp}(\mathbb{I}-\Delta^{qp})}^{-1}\cdot \sum_{\boldsymbol{k}}\ \Delta_{\boldsymbol{k}}^t\cdot R^\dagger \cdot \epsilon(\boldsymbol{k}),
        \label{eq:Veq}
    \end{equation}
    where $V_{BZ}$ is the Brillouin zone volume, and $\Delta_{\boldsymbol{k}} = \langle d^\dagger_{\boldsymbol{k}a}d^\dagga_{\boldsymbol{k}b}\rangle_{qp}$.
    The local potential $\lambda^c$ follows
    \begin{equation}
        \lambda^c_{ab} = -\lambda_{ab}+\frac{\partial}{\partial \Delta^{qp}_{ab}}\mathrm{Tr}\left[R\cdot\sqrt{\Delta^{qp}(\mathbb{I}-\Delta^{qp})}\cdot V\right]+\mathrm{h.c.},
        \label{eq:lceq}
    \end{equation}
    where the derivative acts only on the term on the square root.
    \item Solve the impurity model in Eq.~\ref{eq:Himp} with the previously determined $V$ and $\lambda^c$, and evaluate the impurity model ground state density matrix elements $\langle c^\dagger_{\alpha}d^\dagga_{a}\rangle_{imp}$ and $\langle d^\dagger_{a}d^\dagga_{b}\rangle_{imp}$.
    \item From the solution of the impurity model, the next guess for $R$ and $\lambda$ can be determined. For this, we need to first determine the new quasiparticle density matrix as
    \begin{equation}
        \Delta^{qp}_{ab}=\mathbb{I}-\langle d^\dagger_{a}d^\dagga_{b}\rangle_{imp}.
        \label{eq:DeltaSCF}
    \end{equation}
    From this, the new $R$ parameter can be obtained from
    \begin{equation}
        \sum_a R_{\alpha a} \left[\sqrt{\Delta^{qp}(\mathbb{I}-\Delta^{qp})}\right]_{ab}=\langle c^\dagger_{I\alpha}d^\dagga_{b}\rangle_{imp}.
        \label{eq:RSCF}
    \end{equation}
    Finally, to evaluate a new $\lambda$ there are two possible strategies: determining it through a fitting problem, to enforce that the local one-body density matrix of the new $H_{qp}$ matches the $\Delta^{qp}$ obtained in Eq.~\eqref{eq:DeltaSCF} (cf. Ref.~\cite{Mejuto2023a}), or alternatively to invert Eq.~\eqref{eq:lceq} solving it for $\lambda$ instead of $\lambda_c$ (cf. Ref.~\cite{Mejuto2024}).
    At self-consistency, both aproaches lead to the same solution.
    \item If the new $R$, $\lambda$ match the ones from the previous iteration to some tolerance, the algorithm can be stopped at the stage.
    Otherwise, return to step 1.
\end{enumerate}

It is worth noting that, within the infinite dimensional limit, this self-consistency procedure is equivalent to variationally optimizing a Gutzwiller-like wave function Ansatz for the lattice~\cite{Fabrizio2007,lanata2015}.
Upon convergence, local observables can be directly obtained from the impurity model, whereas non-local observables can be extracted from the quasiparticle Hamiltonian.
In particular, the Green's function can be estimated by projecting the quasiparticle Green's function into the physical orbitals as

\begin{equation}
    G_{\alpha\beta}(\omega,\boldsymbol{k}) = \sum_{ab}R^{\dagga}_{\alpha a}\,\left[\frac{1}{(\omega+i0^+)\mathbb{I} - H_{qp}(\boldsymbol{k})}\right]_{ab}\,R^{\dagger}_{b\beta}.
    \label{eq:gGut_GF}
\end{equation}

Essentially, this allows to map the spectrum of the correlated model into a band structure theory in a larger Hilbert space.
From this Green's function we extract the band structures shown in the main paper.
The ground state energy per site of the model, which we use to compare phase stability invoking the underlying variationality of the method, can be reconstructed from the ground state energy of the impurity model and the density matrix of the quasiparticle Hamiltonian as 

\begin{equation}
    E_0^G/N = E_{imp} + \frac{1}{V_{BZ}}\sum_{\boldsymbol{k}}\sum_{ab} \left(\sum_{\alpha\beta}\, R^{\dagger}_{a\alpha}\, \epsilon_{\alpha\beta}(\boldsymbol{k}) \,R^{\dagga}_{\beta b}\right)\ \Delta^{qp}_{a b}(\boldsymbol{k}).
    \label{eq:GutzEnergy}
\end{equation}

%% file: sections/S4.tex
\section{Energetic stabiliy of altermagnetism over ferromagnetism}

We present a simple mean-field argument that shows that the local altermangetic phase we present is energetically more stable than a ferromagnetic state, where the spin on the two orbitals are aligned.

Consider the Hamiltonian in Eq. \eqref{eq:model}, where the $t_4$ NNN inter-orbital hopping is set to zero. This correspond to two decoupled square-lattice Hubbard models. In the vicinity of the van Hove singularity the single Hubbard models will order ferromagnetically. Consider ordering in the $\sigma_z$ direction, at mean-field level the models will now have the following Hamiltonian:
    
\begin{equation}
    \hat{H}_{2FM} = \sum_{\mathbf{k},s} 
    \begin{pmatrix} 
    c^\dagger_{\mathbf{k} \textit{x} s} & c^\dagger_{\mathbf{k} \textit{y} s} 
    \end{pmatrix}  
    \begin{pmatrix}  
    \epsilon_x(\mathbf{k}) +(-1)^s m_x & 0 \\
    0  & \epsilon_{y} (\mathbf{k})  +(-1)^s m_y  
    \end{pmatrix}
    \begin{pmatrix}
    c_{\mathbf{k}x}\\ c_{\mathbf{k}y}
    \end{pmatrix}    
    \label{eq:SI-model}
\end{equation}
with $m_x , m_y = \pm m$.
Consider the ferromagnetic case, namely $m_x=m_y=m$. We call $\epsilon^+ = \frac{\epsilon_x(\mathbf{k})+\epsilon_y(\mathbf{k})}{2}$ , $\epsilon^- = \frac{\epsilon_x(\mathbf{k})-\epsilon_y(\mathbf{k})}{2}$ and $h(\mathbf{k}$ the inter-orbital hopping in reciprocal space.
Adding back the inter-orbital hopping would result for each $\mathbf{k}$  in the following 4 eigenvalues:
\begin{align*}
\varepsilon^{FM}_1 =& \epsilon^+ + m + \sqrt{ (\epsilon^-)^2+h^2)} \\
\varepsilon^{FM}_2 =& \epsilon^+ + m - \sqrt{ (\epsilon^-)^2+h^2)} \\
\varepsilon^{FM}_3 =& \epsilon^+ - m + \sqrt{ (\epsilon^-)^2+h^2)} \\
\varepsilon^{FM}_4 =& \epsilon^+ - m - \sqrt{ (\epsilon^-)^2+h^2)} \\
\end{align*}
 The two lowest ones will be either $\varepsilon^{FM}_2 , \varepsilon^{FM}_4$ or $\varepsilon^{FM}_3 , \varepsilon^{FM}_4$.

For the altermagnetic case instead the magnetisation are opposite ($m_x=-m_y=m$) and the eigenvalues would be:

\begin{align*}
\varepsilon^{ALM}_1 =& \epsilon^+ + \sqrt{ (\epsilon^- +m)^2+h^2)} \\
\varepsilon^{ALM}_2 =& \epsilon^+ + \sqrt{ (\epsilon^- -m)^2+h^2)} \\
\varepsilon^{ALM}_3 =& \epsilon^+ - \sqrt{ (\epsilon^- +m)^2+h^2)} \\
\varepsilon^{ALM}_4 =& \epsilon^+ - \sqrt{ (\epsilon^- -m)^2+h^2)} \\
\end{align*}

with $\varepsilon^{ALM}_3 , \varepsilon^{ALM}_4$ always being the smallest ones. It is easy to verify that the sum of eigenvalues coming from the latter ordering are always smaller than the ferromagnetic ones, therefore the uniform altermagnetic ordering is favored over the ferromagnetic phase.
We have to consider two cases.
\paragraph{First case:} $m > \sqrt{ (\varepsilon^-)^2 + h^2 }$
\\
In this case we need to prove that
\begin{equation}
    \sqrt{ (\epsilon^- -m)^2+h^2)}+\sqrt{ (\epsilon^- +m)^2+h^2)} > 2m
\end{equation}
We first rewrite this as:
\begin{equation}
    m[ \sqrt{ 1+(\frac{h}{m})^2 + (\frac{\varepsilon^-}{m})^2+2\frac{\varepsilon}{m} } + \sqrt{ 1+(\frac{h}{m})^2 + (\frac{\varepsilon^-}{m})^2-2\frac{\varepsilon}{m} }  ] > 2m
\end{equation}
Introducing $x=\frac{\varepsilon}{m}$ and $y=\frac{h}{m}$ it is easy to prove that
\begin{equation}
    \sqrt{1+y^2+x^2+2x}+\sqrt{1+y^2+x^2-2x} -2 > 0
\end{equation}
The left-hand side is zero only when $y=0$, that correspond to $h=0$, a sub-set of zero measure in the Brillouin Zone.

\paragraph{Second case:} $m < \sqrt{ (\varepsilon^-)^2 + h^2 }$

In this case we need to prove that
\begin{equation}
    \sqrt{ (\epsilon^- -m)^2+h^2)}+\sqrt{ (\epsilon^- +m)^2+h^2)} > 2\sqrt{h^2+\varepsilon^2}
\end{equation}
We rewrite the equation as
\begin{equation}
    \sqrt{ (\epsilon^-)^2+h^2)} \Big[ \sqrt{1+\frac{m^2}{h^2+(\epsilon^-)^2}+\frac{2m\varepsilon}{h^2+(\epsilon^-)^2}}+\sqrt{1+\frac{m^2}{h^2+(\epsilon^-)^2}-\frac{2m\varepsilon}{h^2+(\epsilon^-)^2}} \Big] > 2\sqrt{h^2+\varepsilon^2} \label{eq:tosquare}
\end{equation}

We then call $y=\frac{m}{ \sqrt{h^2+(\varepsilon^-)^2} }$ and $x=\frac{2 \varepsilon m}{h^2+(\varepsilon^-)^2}$.

Clearly when $m>2 \varepsilon$ both terms in the square parentesis on the left are greater than $1$ and the condition is easily realised. If $m=2 \varepsilon$ the left-  and righ-and-sides of the equation are equal but this correspond again to a zero measure set of the BZ.

When $m<2 \varepsilon$ instead we can square Eq. \eqref{eq:tosquare} and obtain:
\begin{equation}
    2\frac{m^2}{h^2+(\varepsilon)^2} + 2\sqrt{ \big( 1+ \frac{m^2}{h^2+(\varepsilon)^2})^2 - (\frac{2 \varepsilon m}{h^2+(\varepsilon)^2} )^2 }>0,
\end{equation}

which is always realised. Therefore the altermagnetic phase is always energetically favorable with respect to the ferromagnetic one due to the presence of inter-orbital hoppings.

%% file: sections/S5.tex
\section{Existence of Dirac points}

The existence of Dirac points is linked to the magnitude of the magnetic gap. Consider in particular a intra-orbital magnetic gap $\Delta_{mag}$ that is opposite for the two orbitals as in our altermagnetic state.
The low energy Hamiltonian in reciprocal space will read:
\begin{equation}
    \hat{H}_{\mathbf{k}, s} = 
    \begin{pmatrix} c^\dagger_{\mathbf{k} \textit{x} s} & c^\dagger_{\mathbf{k} \textit{y} s} \end{pmatrix}  
    \begin{pmatrix}  \ \epsilon_x(\mathbf{k}) +(-1)^s \Delta_{mag} & \epsilon_{xy}(\mathbf{k}) \\ \epsilon_{xy}(\mathbf{k})  & \ \epsilon_{y} (\mathbf{k}) -(-1)^s \Delta_{mag}\end{pmatrix}
    \begin{pmatrix} c_{\mathbf{k} \textit{x} s} \\ c_{\mathbf{k} \textit{y} s}     
    \end{pmatrix}  
\end{equation}

that has eigenvalues:
\begin{equation}
    \varepsilon_{\pm}(\mathbf{k}) = \epsilon^+(\mathbf{k}) \pm \sqrt{[\epsilon^-(\mathbf{k}) +\Delta_{mag}]^2 + [\epsilon_{xy}(\mathbf{k})]^2   }
\end{equation}
with
\begin{align}
    \epsilon^{\pm}(\mathbf{k})  = \frac{\epsilon_x(\mathbf{k}) \pm\epsilon_y(\mathbf{k}) }{2} 
\end{align}

Dirac points appear for k-points that cancel the square root terms in the eigenvalues, that is
\begin{align}
    \epsilon^-(\mathbf{k}) +\Delta_{mag} =0 \\
    \epsilon_{xy}(\mathbf{k}) = 0
\end{align}
The former cannot be verified for
\begin{equation}
    \frac{\Delta_{mag}}{|t_1-t_2|} >1
\end{equation}